# Multi-diseases detection with memristive system on chip

*Zihan Wang*[1,3]  *Daniel W. Yang*[1,3]  *Zerui Liu*[1,3]  *Evan Yan*[1]  *Heming Sun*[1]  *Ning Ge*[2]  *Miao Hu*[2]  *Wei Wu*[1,*]

[1] Ming Hsieh Department of Electrical Engineering, University of Southern California, Los Angeles, CA 90089, USA.
[2] TetraMem Inc., Fremont, CA, USA.
[3] Zihan Wang, Daniel W. Yang and Zerui Liu contributed equally
*Email: wu.w@usc.edu

This study presents the first implementation of multilayer neural networks on a memristor/CMOS integrated system-on-chip (SoC) to simultaneously detect multiple diseases. To overcome limitations in medical data, generative AI techniques are used to enhance the dataset, improving the classifier's robustness and diversity. The system achieves notable performance with low latency, high accuracy (91.82%), and energy efficiency, facilitated by end-to-end execution on a memristor-based SoC with ten 256x256 crossbar arrays and an integrated on-chip processor. This research showcases the transformative potential of memristive in-memory computing hardware in accelerating machine learning applications for medical diagnostics.

## 1. Introduction

Blood tests are integral to diagnosing various diseases and conditions by analyzing substances in the blood, such as cells, chemicals, and proteins[1,2]. Abnormal levels of these components can signal potential health risks, aiding in diagnosing a broad spectrum of illnesses, infections, and diseases[3-5]. However, traditional blood test methods are hampered by significant drawbacks, including prolonged processing time and limited accuracy[1-4]. These methods typically require at least 15 minutes to separate serum from the blood, isolate target substances, and generate results[5-8]. Moreover, their low sensitivity often relegates them to supplementary roles rather than primary diagnostic tools[9-11].

Raman spectroscopy, based on the inelastic scattering of photons, offers a promising alternative for blood test, enabling label-free biomarker detection[12-20]. When integrated with a nanofinger platform, Raman signals, which are naturally weak, can be significantly enhanced for detection[21-25]. Additionally, neural networks, widely recognized for their proficiency in classification tasks, can be crucial in distinguishing signals between healthy individuals and patients[26-31].

Memristor-based neural networks overcome the Von-Neumann bottleneck inherent in traditional digital computing systems by co-locating data storage and computation, thus eliminating the need for frequent data transfers[32]. By directly utilizing Ohm's law for multiplication and Kirchhoff's law for summation, memristor crossbar arrays substantially improve energy efficiency and throughput in deploying deep neural networks [33,34].

Previous studies have demonstrated the potential of memristor-based systems in disease detection, such as in heart attack prevention and breast cancer classification[29,35]. However, these studies reveal several limitations. Primarily, these systems were either simulated or implemented on bulky experimental setups (instead of a fully integrated SoC), limiting their deployment in portable end-user systems and reducing practicality for doctors and patients[36-38]. Additionally, these systems were designed to detect only a single disease, mainly constrained by hardware computing capacity[39-41]. Lastly, due to the high costs and privacy concerns of data collection, deep neural networks were trained on small datasets, leading to overfitting and diminished model reliability[42-45].

In this study, we present a compact, robust, and efficient disease detection system capable of the early prevention of Acute Myocardial Infarction (AMI) and the detection of liver cancer simultaneously. This system is implemented on a memristor/CMOS fully integrated SoC, paving the way for a portable and multifunctional diagnostic system. The SoC consists of ten 256x256 memristor arrays, offering abundant computational capability to support a complex multi-disease recognition network. Furthermore, generative AI techniques were employed to augment the collected dataset, improving model accuracy and robustness against hardware non-idealities[46,47]. The system offers portability, low latency, high accuracy, and energy efficiency, showcasing the potential of memristive in-memory computing hardware to accelerate machine learning and tackle challenges in medical analysis.

## 2. The multi-disease detection system

To develop an efficient and accurate multi-disease detection system, we propose a design that seamlessly integrates both hardware and software components. As illustrated in **Figure 1a**, the data collection process is facilitated by a nanofiber platform that enhances Raman signals. The collected data are subsequently processed using a stable diffusion model, which generates high-quality data by modulating key features from the Raman spectra. Leveraging this augmented dataset, we train a robust classifier capable of detecting liver cancer and heart attacks simultaneously with high accuracy. This classifier is then transferred to our memristive-based SoC platform, enabling rapid and precise detection. With these components integrated, the system setup is completed and ready for daily detection applications. In such daily applications, Raman signals can be directly extracted from blood samples and fed into the SoC, which delivers diagnostic results with minimal delay, facilitating timely medical interventions and enhancing patient outcomes.

### 2.1 Data Collection

**Figure 1b** presents a schematic representation of the nanofinger platform, which is designed to capture biomarkers and enhance Raman signals. The nanofinger platform operates by the collapsing of adjacent fingers and then forming a finger-antibody-finger sandwich structure, driven by capillary forces after antibody soaking and air drying. Antibodies are covalently bonded to gold nanoparticles via thiols, allowing receptors to detect target biomarkers within complex blood components. These biomarkers are immobilized in the gap between nanofingers, thereby maximizing electromagnetic field enhancement through gap-plasmon resonance. This design achieves high selectivity with a simple surface treatment. Subsequently, the enhanced Raman signals from the captured biomarkers are subsequently collected.

### 2.2 Data Generation

To generate high-fidelity Raman signals that contribute to a robust and reliable deep neural network (DNN) model, this study employs a conditional diffusion model integrated with cross-attention and self-attention layers (**Figure 1c-d**). This approach offers several advantages: (i) The self-attention mechanism effectively captures global features of long sequences of Raman signals, filtering out irrelevant local fluctuations and providing a comprehensive overview of the global trend. (ii) By integrating conditioning features into the diffusion model (DM) through cross-attention layers, the inductive bias of DM can significantly modulate the conditioning context information and the generating targets. This integration allows for guided and controlled signal generation, enhancing the overall performance and reliability of the DNN model.

#### 2.2.1 Diffusion Model

The diffusion model is a generative model designed to learn the distribution $p(x)$ by modeling the data generation process as a series of denoising steps that gradually transform noise into a structured data. During the Forward Diffusion Process, a series of Gaussian noises are incrementally added to the clean generating target $x_0$. This creates a sequence of increasingly noisy images $x_{t1}, x_{t2}, \ldots, x_{tn}$, where $t_i$ is uniformly sampled from $[0, T]$. Mathematically, this process is described as the equation below:

$$q(x_t \mid x_{t-1}) = \mathcal{N}(x_t; \sqrt{1 - \beta_t} x_{t-1}, \beta_t I) \tag{1}$$

where $\beta_t$ is a small positive constant that controls the variance of the added noise at each step. The backbone U-Net architecture aims to predict the denoised variant at each time step $t$, which can be represented as

$$L_{\text{DM}} = E_{x_0, \epsilon, t}[||\epsilon - \epsilon_\theta(x_t, t)||^2] \tag{2}$$

where $\epsilon$ is the noise added to the data sample, $\epsilon_\theta(x_t, t)$ is the neural network's prediction of the noise at timestep $t$, $E$ denotes the expectation over the data distribution and the noise distribution. In the sampling process, pure Gaussian noise is input into the U-Net and is gradually denoised, applying the learned reverse diffusion steps iteratively to obtain a clean data. As a result, the target $x_0$ is generated as

$$p_\theta(x_{t-1} \mid x_t) = \mathcal{N}(x_{t-1}; \mu_\theta(x_t, t), \Sigma_\theta(x_t, t)) \tag{3}$$

Where $\mu_\theta(x_t, t)$ is the predicted mean for the reverse process and is given by the neural network with parameters $\theta$, and $\Sigma_\theta(x_t, t)$ is the predicted variance for the reverse process, often simplified as a fixed variance or learned by the model.

### 2.2.2 Conditioning Mechanisms

The traditional U-Net structure, based on convolutional layers, is limited in handling sequential data and capturing long-term dependencies, and struggles to aggregate conditioning information effectively. Recent advancements have demonstrated that integrating transformers into the U-Net backbone of diffusion models (see supporting information, **Extended Figure 1**) significantly enhances the performance and enables various token-based conditioning mechanisms. Our study applies this architecture to the field of medical analysis, resulting in the generation of a substantial amount of high-quality data.

Specifically, the conditioning information is obtained by Principle Component Analysis (PCA), based on the observation that PCA highlights the primary features of a specific Raman signal. To effectively modulate the conditioning pair, a cross-attention block is designed as

$$Attention(\mathbf{Q}, \mathbf{K}, \mathbf{V}) = softmax\left(\frac{\mathbf{Q}\mathbf{K}^T}{\sqrt{d_k}}\right)\mathbf{V} \tag{4}$$

with

$$Q = \mathbf{X}\mathbf{W}_Q, \quad K = \mathbf{C}\mathbf{W}_K, \quad V = \mathbf{C}\mathbf{W}_V \tag{5}$$

where $\mathbf{C}$ is the matrix of conditioning vectors, and $\mathbf{W}_Q, \mathbf{W}_K, \mathbf{W}_V \in R^{d \times d_k}$ are learnable projection matrices, $x_i \in R^d$ denotes a sequence of input vectors and $c_i \in R^d$ represents conditioning vectors. Similarly, self-attention is integrated to capture the global trend of long Raman signals as **equation 4**, with

$$Q = \mathbf{X}\mathbf{W}_Q, \quad K = \mathbf{X}\mathbf{W}_K, \quad V = \mathbf{X}\mathbf{W}_V \tag{6}$$

With the conditioning information embedded, the diffusion model aims to learn a conditional distribution $P(x|c)$. Accordingly, it optimizes the loss function as

$$\mathcal{L}_{CDM} = E_{t,x_0,\epsilon,c}[||\epsilon - \epsilon_\theta(x_t, t, c)||^2] \quad (7)$$

Here $x_t$ is the noisy version of $x_0$ at time step $t$, generated by adding Gaussian noise $\epsilon$. $\epsilon_\theta(x_t, t, c)$ is the U-Net with attention mechanisms that predict the noise $\epsilon$, conditioned on $x_t$, the time step $t$, and the conditioning information $c$. $\epsilon$ is sampled from a standard normal distribution $\mathcal{N}(0, \mathbf{I})$.

## 2.3 Classification

DNN is employed to model the relationship between Raman signals and diagnostic outcomes. In prior work, although a small DNN achieved a decent accuracy, its performance was constrained by the bulky size but small capacity of the system and faced challenges with multi-disease classification tasks. In this study, as displayed in **Figure 1e**, a deeper multilayer perceptron (MLP) with integrated ReLU activation functions is designed. As the length of the Raman signal sequence exceeds the capacity of the SoC crossbar, preprocessing with PCA is used to down sample the input first. The classifier's performance is optimized using cross-entropy loss.

Given that programming errors are a significant factor degrading performance in crossbar arrays, it is important to make the model robust to such variations. An intuitive approach is to minimize the absolute values of the weights, ensuring that after quantization, the mathematical discrepancy between the ideal and experimental weights is minimized. This can be achieved by adding an $L_2$ regularization term to the cross-entropy loss function, as shown in the following equation:

$$\mathcal{L} = \mathcal{L}_{cross\text{-}entropy} + \lambda ||\mathbf{W}||_2^2 \quad (8)$$

where $\mathcal{L}_{cross\text{-}entropy}$ represents the standard cross-entropy loss, $\lambda$ is a regularization parameter that controls the trade-off between the cross-entropy loss and the $L_2$ regularization and $||\mathbf{W}||_2^2$ denotes the $L_2$ norm of the weight matrix $\mathbf{W}$.

## 2.4 Hardware Transfer

The operation of a linear layer can be represented by the following equation:

$$\mathbf{y} = \mathbf{W}x + b \quad (9)$$

where $y$ is the output vector, $\mathbf{W}$ is the weight matrix, $x$ is the input vector, and $b$ is the bias vector. Given that the memristive-SoC is designed to accelerate vector-matrix multiplication (VMM), this equation can be reformulated as a fully matrix-based multiplication:

$$\mathbf{Y} = \mathbf{W}'\mathbf{X}' \quad (\mathbf{10})$$

where $\mathbf{W}' = [W \quad b]$, $\mathbf{X}' = \begin{bmatrix} X \\ 1 \end{bmatrix}$. Upon quantizing the weights and inputs into 8-bit integers (see supporting information, **Extended Figure 3**), the four-layer multi-disease classifier can be mapped as illustrated in the accompanying **figure 1f** and **1g**. Considering that our SoC supports only unsigned 8-bit integer inputs, while the PCA results include negative values, the operation of the first linear layer is split into two matrix multiplications:

$$\begin{aligned}
\mathbf{X}^+ &= \max(\mathbf{X}, 0), \\
\mathbf{X}^- &= \max(-\mathbf{X}, 0), \\
\mathbf{Y} &= \mathbf{W}^+\mathbf{X}^+ - \mathbf{W}^-\mathbf{X}^-
\end{aligned} \quad (11)$$

By using this approach, the precision of the first layer is effectively increased to 9 bits, at the expense of an additional VMM operation for the SoC. This trade-off is particularly beneficial when considering the wide dynamic range of real-valued mathematical inputs.

## 3. MX100 SoC

The development of memristor-based IMC is advancing rapidly across applications, workflows, algorithms, and data augmentation. Historically, the implementations of memristor-based systems have been constrained by bulky printed circuit boards (PCBs) or probe cards, which support only a small number of memristor cells. These configurations not only offer restricted computational capabilities but also occupy considerable physical space. In contrast, modern artificial intelligence (AI) models incorporate a vast number of parameters, necessitate high levels of parallelism to achieve adequate throughput, and require highly integrated hardware platforms for seamlessly embedding into compact end systems.

The TetraMem® MX100 SoC, an analog IMC accelerator built with the 65-nm technology node, represents the first fully integrated memristor/CMOS computing chip featuring 8-bit memristors (**Figure 2a**). As illustrated in **Figures 2b and 2c**, the chip comprises ten neural processing units (NPUs), specifically optimized for VMM and depth-wise operations, enabling massive parallelism in processing. The chip is also equipped with a RISC-V on-chip central processing unit (CPU) to execute on-chip program, a direct memory access (DMA) engine to facilitate rapid on-chip data transfer, a 512 KB of SRAM dedicated to both instruction and data storage, and a 1 MB of system SRAM for NPU buffer data storage.

Upon reset, the system initiates the execution of the on-chip program stored in RAM by the RISC-V core. The input matrix is transferred through the I/O peripheral, after which the vector matrix computation is accelerated by the ten NPUs utilizing the in-memory computing architecture, with one NPU being specifically designed to optimize depthwise convolution operations. Finally, the result of the dot product computation is output through the I/O peripheral.

NPUs, the core of the IMC architecture, are constructed with 256x256 8-bit computing memristor arrays. As illustrated in **Figure 2d**, each memristor is fabricated with an accompanying transistor to form a 1T1R cell. Each NPU comprises 256 wordlines, driven by an 8-bit Digital-to-Analog Converter (DAC), and 256 select lines connected to the gate terminals of the transistors, which are controlled by a selector control mechanism. Furthermore, the architecture includes 256 bitlines, through which the accumulated current is digitized by an 8-bit Analog-to-Digital Converter (ADC) to generate the output (see supporting information, **Extended Figure 4**).

The NPUs on the chip exhibit highly reproducible and uniform multi-level conductance states, as demonstrated by the test results presented in **Figure 2e**. This figure shows the Cumulative Distribution Function (CDF) for 65,536 memristors distributed across 256 distinct conductance states, with the curves displaying clear separation and near-vertical alignment, though some instances of stuck-on and stuck-off cells are still observed. The conductance states were achieved using identical SET and RESET pulse trains, each with a pulse width of 50 ns, employed in closed-loop programming operations.

# 4. Experimental

Leveraging the SoC and an enlarged dataset generated through data augmentation, the developed system exhibits high accuracy, robustness, and throughput. This is empirically validated through the following steps: First, we present the signals generated via the conditional diffusion process and demonstrate the improved accuracy attributable to data augmentation. Next, we showcase a robust hardware demonstration, where the accuracy matches that of the software implementation while operating with reduced energy achieved consumption.

## 4.1 Conditioning Diffusion Model

This section analyzes the performance of the conditional diffusion model with cross-attention layers. **Figure 3a** illustrates the denoising process of a DM applied to Raman signals over multiple time steps. With t decreasing, the signal is turning from a noisy state (at earlier steps) to a cleaner, more defined form (at later steps), which depict gradual reductions in noise. The black line represents the final generated Raman signal after the completion of the denoising process, highlighting the effectiveness of the DM in reconstructing the original signal from noisy data.

The analysis of the guided generation process by aggregating context information in DM is shown in **Figure 3b** (see supporting information **Extended Figure 2** for more examples), where each column corresponds to a distinct sample. The black plots in the bottom row of each column depict the original data, serving as a reference for comparison. The colored plots in the columns above the original data represent the generated samples guided by the context information, which is specifically the PCA results. It can be observed that the generated samples closely resemble the original data, while at the same time some diversity is still maintained, thus validating the model's ability to capture and utilize contextual information for generating high-fidelity samples and potentially leading to more robust and generalized training for generating high-fidelity samples and potentially leading to more robust and generalized model training.

**figure 3c** illustrates the distribution of samples across three categories—Healthy individuals, Heart Attack patients, and Liver Cancer patients—before and after data augmentation. In the original dataset, each category contains 431, 385, and 212 samples, respectively. Data augmentation increases these numbers to 2012, 1540, and 852, respectively.

The testing accuracies with and without data augmentation are shown in **Figure 3d**. In the scenario without augmentation, the dataset was partitioned into training, validation, and testing sets in a ratio of 7:2:1. In the scenario with data augmentation, to validate the robustness of our model under real-world conditions, all the testing samples were from the original dataset, while the rest of the dataset was randomly partitioned into training and validation sets. The final ratio remains approximately 7:2:1. The testing experiment on the original dataset achieved an average accuracy of 87.54%, which serves as the baseline for subsequent experiments. With data augmentation, the performance increased by 6.21% and even 7.79% when additional samples were generated for underrepresented categories, namely through data balancing. Furthermore, the accuracy of individual classes also improved with data augmentation and dataset balancing, with the most significant improvement observed in the liver cancer patient category. This figure demonstrates that high-quality generated data plays a crucial role in enhancing the performance and reliability of the classifier.

The robustness of the model is illustrated in **Figures 3e and 3f**, where the validation loss fluctuates significantly and is higher on average than the training loss. This phenomenon indicates that the model performs poorly on unseen data, primarily due to the limited size of the training set, which leads to overfitting and the memorization of noise rather than generalization from the data. However, following data augmentation, the training loss quickly stabilizes and becomes lower than the validation loss. This improvement occurs because the model learns to capture the underlying patterns in the expanded dataset rather than the noise. Consequently, while the training loss is higher because the model does not fit the noise as closely, the validation loss is lower, reflecting better predictive performance on unseen data.

## 4.2 SoC

As mentioned in Section 2.4, the SoC utilizes split positive and negative int-8 matrices as inputs, derived from PCA results of 514 real-collected samples, each comprising 128 data points. Following processing by the 8-bit DAC, the int-8 values are converted to voltages. These voltages are then directly applied to the wordlines, as illustrated in **Figures 4a and 4b**.

The weights were trained using a quantization-aware training method and subsequently converted to int8 values to align with the 256 conductance states of the memristor device. As detailed in Section 2.4, the weight mapping is depicted in **Figures 4c and 4d**. The primary sources of programming error include device variance, and devices that are stuck-on or stuck-off. To address the issue of real conductance values exceeding the lower and upper bounds of the normalized weights while the reading results remain within the [0, 255] range, we normalized the quantized weights to the range [50, 200].

The output features of the first layer, normalized to math values, are illustrated in **Figure 4e and 4g**(see supporting information **Extended Figure 5** for the other 3 layers). The figures display results for 514 testing samples, each with an output feature vector of 240 dimensions. The difference between the hardware experimental results and the simulated results is minimal. The minor discrepancies are primarily attributed to programming errors.

**Figure 4h** compares the accuracy of the multi-disease recognition model across three different environments: PyTorch, the model post-quantization, and the SoC implementation (see supporting information, **Extended Figure 6** for more experiments). Key observations from this comparison include: (i) The accuracy of the quantized model, where both input and weights are quantized to int-8, and the pure SoC implementation exhibit slight decreases compared to that of the pure software baseline, with reductions of 0.59% and 3.51%, respectively. (ii) The percentage of healthy individual predictions is most vulnerable to quantization and hardware non-ideality, whereas the prediction accuracy for heart attack and liver cancer patients remains relatively stable or even increases by 0.05% compared to the baseline. The inaccuracy tends to be more in false positive than false negative, which is in general less harmful. These results imply that our model maintains high accuracy for critical disease predictions, which could provide valuable assistance to doctors by alerting them to potential patient conditions in real-world application scenarios.

## 5. Conclusion

We have developed a robust multi-disease recognition system based on a memristor SoC. This state-of-the-art hardware platform enables the deployment of the DNN model in a portable, prompt, and efficient manner. The accuracy achieved with the SoC is comparable to that of pure software implementation, with a small discrepancy attributed to factors such as device variance, non-idealities in the ADCs, and precision loss in weights and inputs. Algorithm and hardware co-design can be employed to mitigate these issues. For instance, we utilized Diffusion Probability models to synthesize high-fidelity samples, enhancing the model's resilience to device variance and significantly improving classification accuracy. Our results represent a significant advancement in healthcare technology, made possible by memristive in-memory computing systems.

# Acknowledgements

I would like to express my deepest gratitude to all those who have supported me throughout this research project. I would like to thank my colleagues, Wenhao Song, Tong Wang and Ruoyu Zhao, for their insightful comments and constructive criticism. I am also grateful to University of Southern California and TetraMem Inc. for providing access to their facilities and resources.

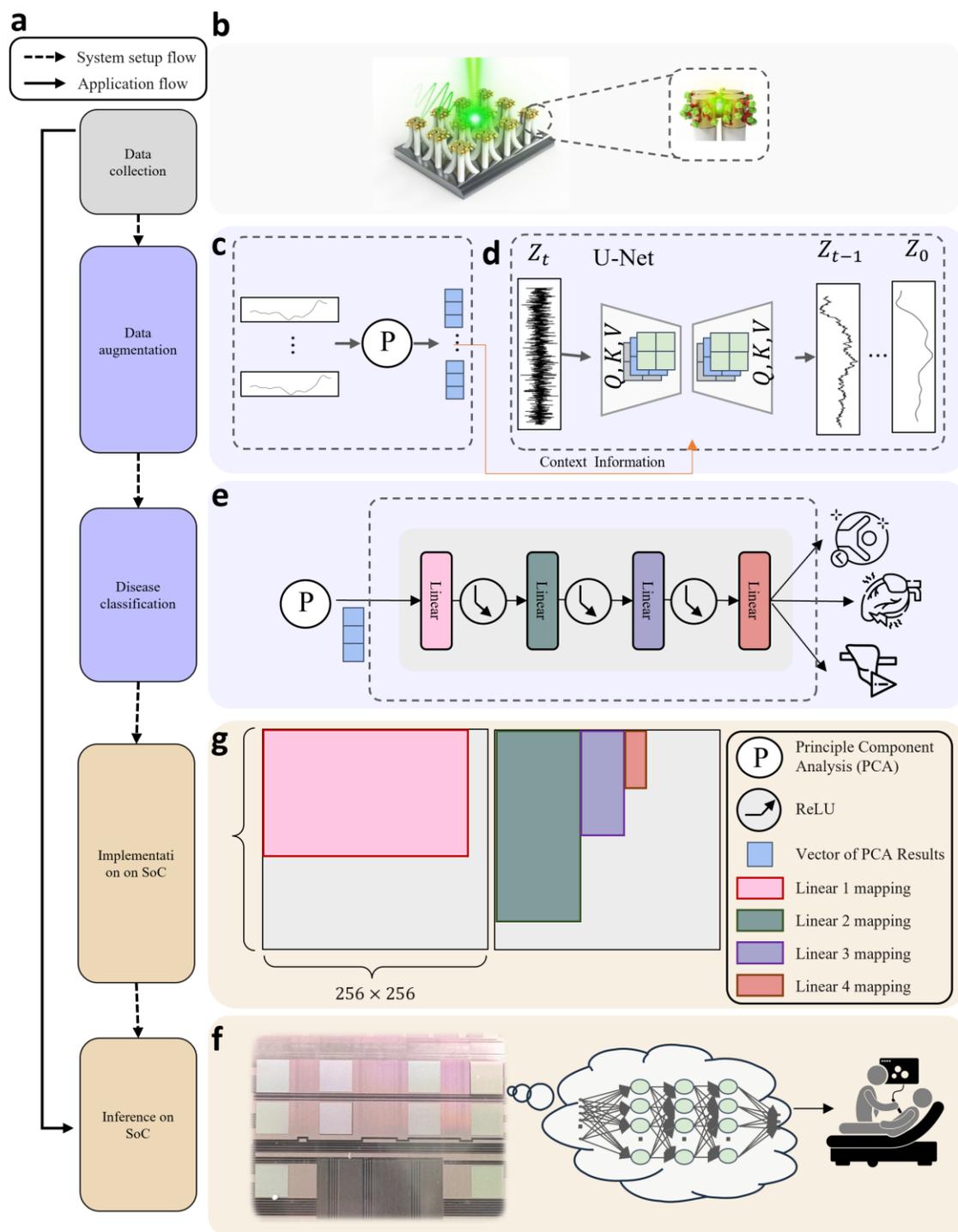

Figure 1: **Disease Detection System. a,** Flowchart of the disease detection system. Data augmentation and classification are performed in software, while weight implementation and inference are executed on the SoC. **b,** Nano-finger platform used to enhance Raman spectroscopy. The left side shows twelve collapsed nanofingers, with hotspots magnified on the right side. **c-d,** Diagram of the reverse process of conditional DM. PCA results are input as context information. Pure Gaussian noise is processed by the attention-embedded U-Net and gradually denoised. **e,** The MLP structure serving as a multi-disease classifier. **e-f,** Weight mapping of four different layers of the classifier: Colors correspond to those in the MLP structure figure.

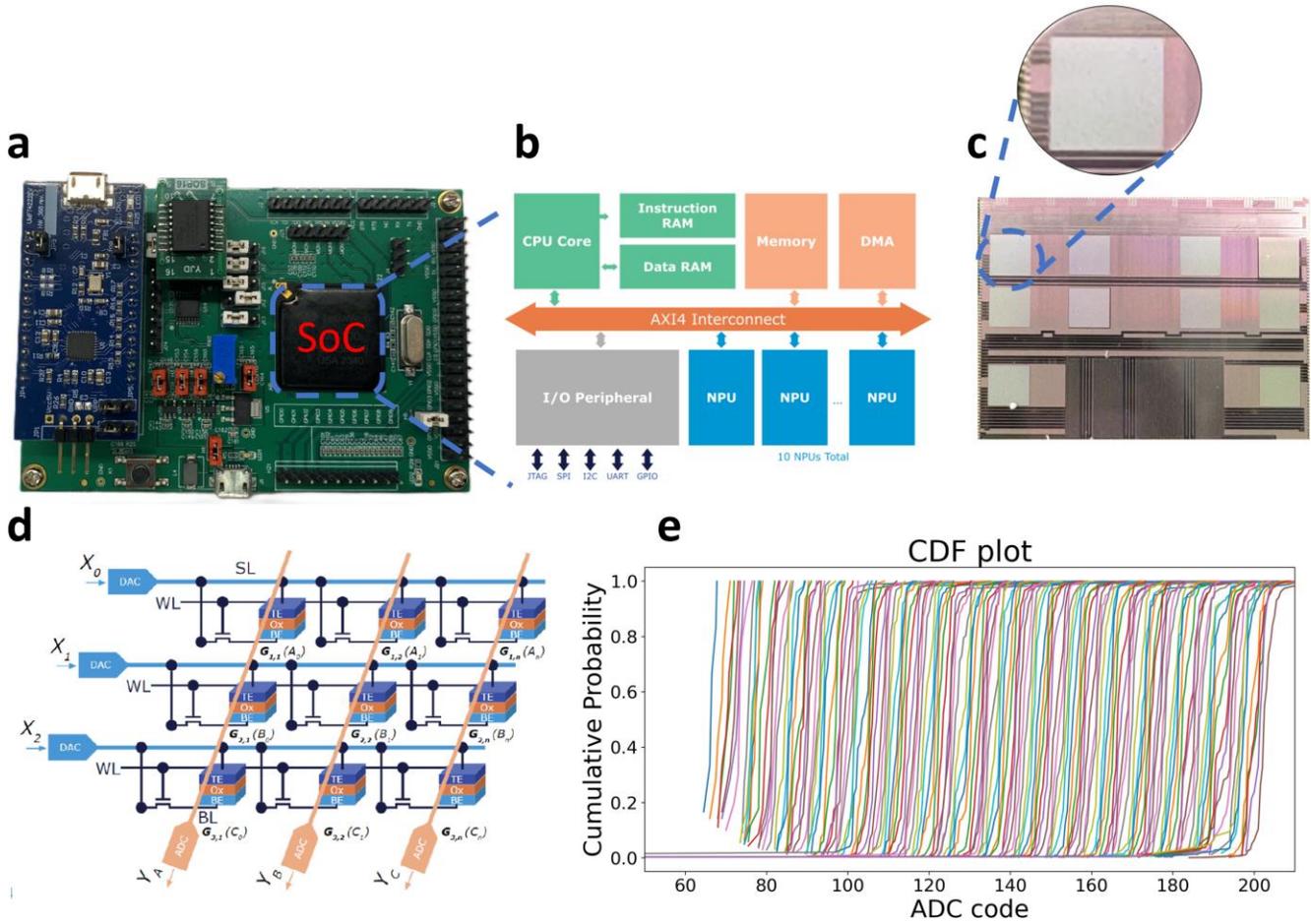

Figure 2: **Reliable Memristor-based SoC System. a,** Photograph of the MX100 SoC evaluation board. **b,** Diagram of the SoC architecture featuring 10 NPUs and one RISC-V CPU integrated. **c,** Photograph of the 10 NPUs, including a zoomed-in view of a sigle NPU. **d,** Acceleration mechanism inside the NPU. By applying Ohm's Law and Kirchhoff's Law, the results of the VMM are obtained by measuring the current of each bitline. **e,** Cumulative Distribution Function (CDF) of 256 reliable conductance levels across 65,536 memristor cells. The conductance levels are directly obtained through 8-bit ADC readout.

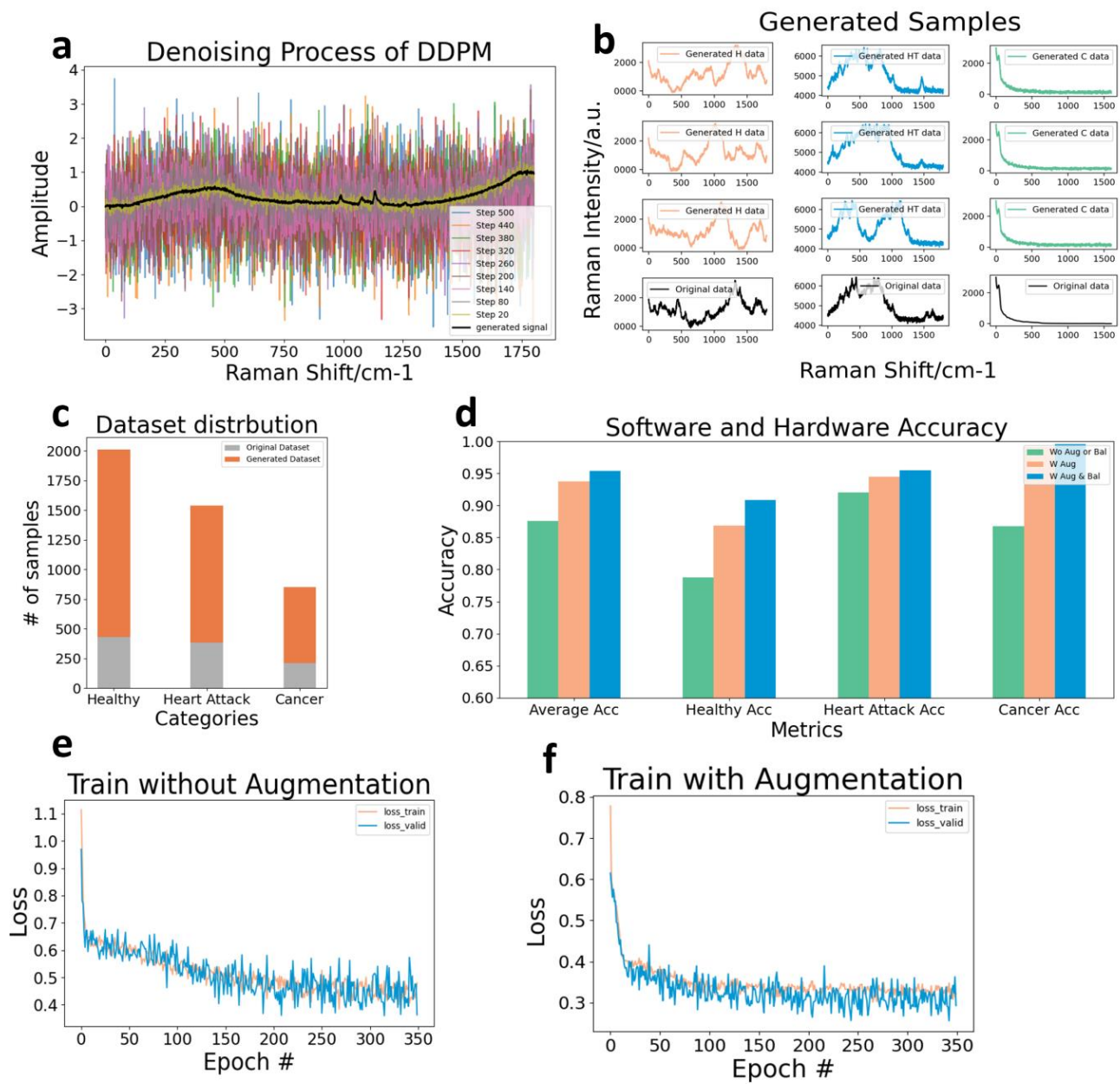

Figure 3: **Data Augmentation strengthens the software classifier. a,** Gradual denoise process of the DM With $T = 500$. The denoised signals are displayed every 60 steps. **b,** Context information-guided generation. The colored signals are synthesized using the bottom black Raman signals as a reference. **c,** Distribution of the multi-disease dataset before and after data augmentation. An additional 3-5 samples are generated using one real Raman spectroscopy as a control. **d,** Accuracy metrics trained on three different datasets: Comparison of performance using all collected signals, equivalently augmented dataset, and dataset-balancing oriented augmented dataset. **e-f,** Illustration of increased robustness before(e) and after(f) data augmentation. Comparison between training loss and validation loss demonstrates improved stability.

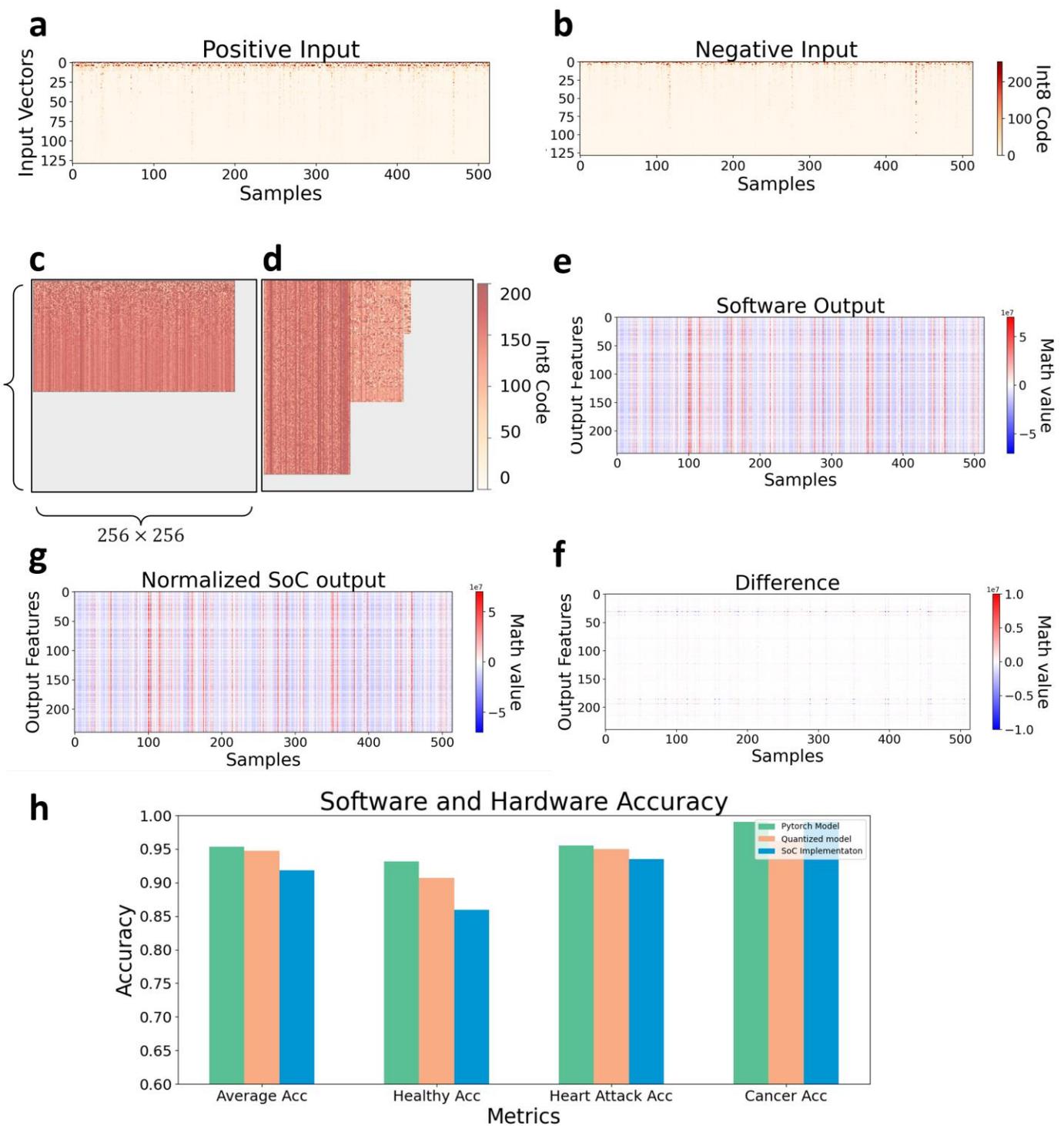

Figure 4: **SoC Achieves Comparable Performance with Pure Software. a-b,** Quantized int-8 input matrices. The input, consisting of 514 vectors each downsampled from 1800 to 128 dimensions, is split into positive and negative components. These are converted to voltages by DACs to drive the wordlines. **d-e,** Partition of quantized int-8 MLP weights. The weights are transferred to two different NPUs. The finetuning criterion is achieving a Root Mean Square Error (RMSE) of less than 5 for the entire weight matrix. **e-g,** Comparison between simulated and hardware VMM results(e-f). The difference between the PyTorch result of the first linear layer and the pure hardware implementation(g) is illustrated. **h,** Accuracy metrics across three different configurations: pure software, simulation with quantized weights and inputs, and the raw SoC implementation.

**Table of Contents**

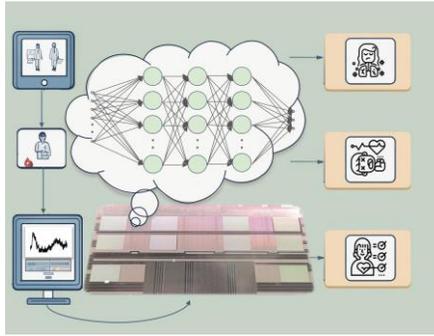

**System Behavior:** A robust disease detection system, which is capable of the early prevention of Acute Myocardial Infarction (AMI) and the detection of liver cancer, was implemented on a memristive SoC. A fully integrated SoC was utilized to ensure the system's portability, low latency, high accuracy, and energy efficiency for medical analysis.